\documentclass[aip, superscriptaddress, preprint]{revtex4-1}
\usepackage{gensymb}
\usepackage{amsmath}
\usepackage{graphicx}% Include figure files
\usepackage{bm}% bold math
\usepackage{hyperref}% add hypertext capabilities
\hypersetup{
            colorlinks=true,
            linkcolor=blue,
            anchorcolor=blue,
            citecolor=blue
            }
\usepackage{changes}

\begin{document}

%\preprint{APS/123-QED}
\title{ Thermal noise induced probability switching in magnetic tunnel junction based on spin-circuit simulation }
% Force line breaks with
%\thanks{A footnote to the article title}%

\author{Shaojie Hu}
%\email%[Authors to whom correspondence should be addressed:\ ]{hushaojie@sztu.edu.cn}
\email[]{hushaojie@sztu.edu.cn} 
\affiliation{College of Integrated Circuits and Optoelectronic Chips, Shenzhen Technology University, 3002 Lantian Road, Pingshan District, Shenzhen Guangdong, China, 518118}

\author{Fupeng Gao}
%\email[]{xxx} 
\affiliation{School of Microelectronics $\&$ State Key Laboratory for Mechanical
Behavior of Materials, Xi'an Jiaotong University, Xi'an 710049, China }

\author{Tengwei Huang}
%\email[]{202201202120@stumail.sztu.edu.cn} 
\affiliation{College of Integrated Circuits and Optoelectronic Chips, Shenzhen Technology University, 3002 Lantian Road, Pingshan District, Shenzhen Guangdong, China, 518118}

\author{Zhizhong Wang}
%\email[]{wangzhizhong@sztu.edu.cn} 
\affiliation{College of Integrated Circuits and Optoelectronic Chips, Shenzhen Technology University, 3002 Lantian Road, Pingshan District, Shenzhen Guangdong, China, 518118}

\author{Hui Li}
%\email[]{lihui3@sztu.edu.cn} 
\affiliation{College of Integrated Circuits and Optoelectronic Chips, Shenzhen Technology University, 3002 Lantian Road, Pingshan District, Shenzhen Guangdong, China, 518118}

\author{Dawei Wang}
%\email[]{dawei.wang@xjtu.edu.cn}
\affiliation{School of Microelectronics $\&$ State Key Laboratory for Mechanical Behavior of Materials, Xi'an Jiaotong University, Xi'an 710049, China }

\date{\today}

\begin{abstract}

The probability switching characteristics in spin transfer torque magnetic tunnel junctions (STT-MTJs) are simulated by considering thermal noise using a spin-circuit module. Thermal noise significantly affects the probability switching for pulse durations exceeding 10 ns, while no probability switching properties are observed for pulses shorter than 1 ns due to the precessional switching. For pulse durations between 1 ns and 10 ns, the occurrence of mixed probability and abrupt switching suggests that thermal noise partially influences the switching properties. These results demonstrate the effectiveness of our simulation model in capturing the MTJ properties under the influence of thermal noise. The spin-circuit module used in this study lays the groundwork for future circuit system designs utilizing MTJ devices, such as true random number generators and neural network computing.  

\end{abstract}
%\keywords{Suggested keywords}

\maketitle
\section{Introduction}
\begin{figure}[h]
	\centering
        \includegraphics[width=6in]{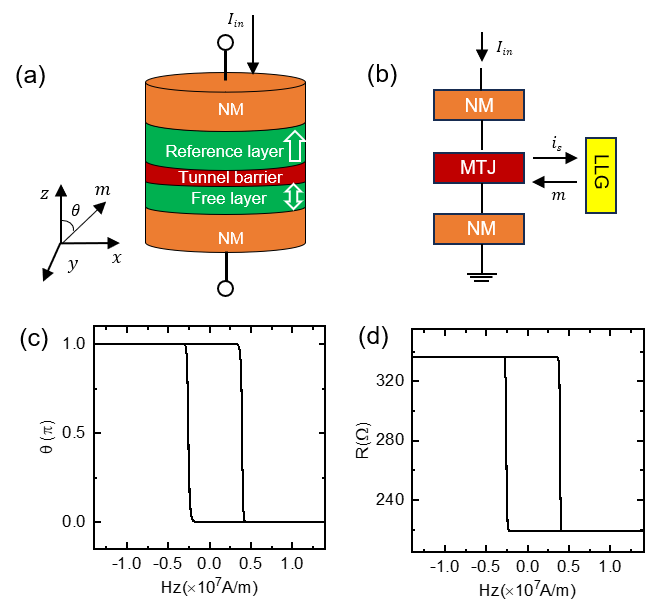}	
	\caption{(a) The structure diagram of STT-MTJ is composed of a non-ferromagnet at both ends and an MTJ, and the $\theta$ Angle is the Angle between the magnetic moment direction of the MTJ free layer and the fixed layer. (b) Spin circuit model of STT-MTJ. (c)The change of $\theta$ Angle with applied magnetic field. (d)The change of MTJ resistance with applied magnetic field.}
 \label{fig1}
\end{figure}

Spin-transfer torque magnetoresistive random access memory (STT-MRAM) is emerging as a promising non-volatile memory technology characterized by low power consumption and high performance. The core structure of STT-MRAM consists of magnetic tunnel junctions (MTJs), typically composed of two ferromagnetic (FM) layers separated by a thin insulating barrier, as illustrated in Fig.\ref{fig1}(a). One FM layer, the free layer (FL), stores binary information via its magnetization orientation, while the other, the reference layer (RL), provides a fixed magnetization reference for reading operations.
The fundamental operating principle of STT-MRAM relies on spin-transfer torque (STT), which induces magnetization switching in nanostructured magnetic materials through spin-polarized currents \cite{SLONCZEWSKI1996L1}. The dynamics of magnetization switching within the FL can be described by the Landau-Lifshitz-Gilbert (LLG) equation \cite{Sun_PRB}.
However, practical performance of spintronic devices, including MTJs, is significantly influenced by magnetization noise \cite{Wang_Xiaobin_2008,PhysRev.130.1677,PhysRevLett.91.146803,Bedau2010}. Particularly, under thermally activated switching conditions, thermal noise becomes the dominant factor due to its pronounced effect in modifying the effective temperature and lowering the energy barrier for magnetization reversal \cite{Wang_APL}. Thermal noise arises from the Brownian motion of electrons in conducting materials, characterized typically by a white noise spectrum and a Gaussian amplitude distribution.
The crucial consequence of thermal noise is the reduction of the energy barrier required for magnetization switching. When the applied current pulse approaches the critical switching current of the MTJ, thermal noise enables probabilistic magnetization switching. Such probabilistic behavior of MTJs facilitates their application in probabilistic computing. However, current simulation techniques primarily employ compact mathematical models, which lack physical-level fidelity. Consequently, these models inadequately account for the influence of pulse duration and circuit density effects on switching behavior, limiting their accuracy and predictive capabilities.

This work utilizes a physics-based simulation framework implemented entirely in SPICE for hybrid spintronic and conventional electronic circuits \cite{Spin_circuit}. This framework offers numerous advantages, including high accuracy, fast simulation times, compatibility with existing MOSFET SPICE models, ease of implementation, and excellent usability. To investigate the probabilistic switching in MTJs induced by thermal noise, we developed a spin circuit comprising an MTJ module, two non-ferromagnetic (NM) modules, and a Landau–Lifshitz–Gilbert (LLG) module, as shown in Fig.\ref{fig1}(b). The MTJ block, structured as a nanopillar, is positioned between two NM modules, each with a radius of 75 nm. The magnetic free layer within the MTJ has a thickness of 3.5 nm. Additional parameters are listed in Table\ref{parameters}. Initially, the magnetization orientations of both the fixed and free magnetic layers are set along the +$\hat{z}$ direction. The magnetization orientation of the free layer is denoted by $\theta$, representing the angle between the +$\hat{z}$ axis and the instantaneous magnetization direction $\hat{m}$.
Fig.~\ref{fig1}(b) also illustrates a schematic diagram of the spin circuit used to simulate spin current transport within the device. The LLG module implements the classical Landau–Lifshitz–Gilbert magnetodynamic equation, incorporating spin dynamics to accurately describe the effects of spin currents and thermal noise on the magnetic moment of the MTJ’s free layer. The employed LLG equation can be expressed as follows:
\begin{equation}
    \left (\frac{1+\alpha^2}{\gamma H_K}\right)\frac{d\hat{m}}{dt}=-\hat{m}\times\Vec{h}_{eff}-\hat{m}\times\hat{m}\times\Vec{h}_{eff}-\alpha\times\hat{m}\times\hat{m}\times \hat{i_s}+\alpha\hat{m}\times \hat{i_s}
\end{equation}
where $\alpha$ is the damping coefficient, $\gamma$ is the gyromagnetic ratio, $\hat{m}$ is the time-dependent magnetization (unitless), $\Vec{h}_{eff}$ is the normalized effective field $\hat{H}_{eff}/H_K$, $\hat{H}_{eff}=\hat{H}_K+\hat{H}_p+\hat{H}_{ext}$, including uniaxial, shape anisotropy as well as the external magnetic, and $i_s$ is the normalized spin-current $\hat{I_s}/H_K$. $\hat{I_s}=\frac{\hbar I_c}{2qM_sV}$ and $I_c$ are the spin-polarized and charge current, which flow to the magnetic free layer. q is the value of electron charge, V is the volume of the magnetic free layer. \par

\begin{table}
    \centering
    \caption{Parameters for the simulation}
    \begin{tabular}{l|c|l}
    \hline
    \textbf {Description} & \textbf{Parameter} & \textbf{Value} \\
    \hline
        Saturation magnetization  & $M_{s}$ & $9\times10^{5}$ A/m\cite{Wang_2015} \\
        Uniaxial anisotropy constant  & $H_{k}$ & $8\times10^{4}$ A/m\cite{Wang_2015}\\
       Damping coefficient  & $\alpha$ & $0.03$\cite{Wang_2015}\\
       Gyromagnetic ratio  &  $\gamma$ & $1.76\times10^{11}\  rad(s\cdot T)^{-1}$\\
       Radius of pillar  & r & $75\times10^{-9}\  m$\cite{Wang2018FieldfreeSO}\\
       Resistivity of NM  & $\rho_{NM}$ & $2.89\times10^{-8}\ \Omega\cdot\  m$\\%一直用的
       Conductivity of junction  & $G_{0}$ & $3.77\times10^{-3} \ s/m$\\%一直用的
       Spin polarization of free layer  & $P_{1}$ & 0.46\\%一直用的
       Spin polarization of fixed layer  & $P_{2}$ & 0.46\\%一直用的
       Spin diffusion length of NM  & $\lambda_{NM}$ & $450\times10^{-9} \ m$\cite{Hu2014EfficientTS}\\
       Spin diffusion length of ferromagnet  & $\lambda_{FM}$ & $2\times10^{-9} 
       \ m$\cite{Cui_2022}\\
        Thickness of fixed layer  & $t_{FM}$ & $3.5\times10^{-9}\  m$\\%一直用的
        Thickness of NM   & $t_{NM}$ & $200\times10^{-9}\  m$\\%一直用的
    \hline
    \end{tabular}
    \label{parameters}
\end{table}

To validate the fundamental operation of the MTJ, simulations were conducted on magnetization switching of the free layer by sweeping an external magnetic field along the z-direction. Initially, both free and fixed layer magnetizations are oriented in the +z direction (i.e., magnetization angle $\theta=0$).
Fig.\ref{fig1}(c) and (d) plot the magnetization angle $\theta$ of the free layer and the MTJ resistance, respectively, as functions of the external magnetic field. As the external magnetic field magnitude decreases from positive values, the magnetization angle $\theta$ remains stable, indicating no switching in the free layer. However, upon reaching a certain negative field value, $\theta$ rapidly shifts from $0$ to $\pi$, indicating the magnetization of the free layer switches from the +z to the -z direction, transitioning the MTJ from a parallel (P) to an antiparallel (AP) state. Correspondingly, the MTJ resistance increases from its initial value of approximately 220 $\Omega$ to 340 $\Omega$, as illustrated in Fig.\ref{fig1}(d).
When sweeping the magnetic field back from negative maximum values toward positive values, the magnetization of the free layer returns to the +z direction at a specific positive magnetic field, and the MTJ resistance reverts to its initial lower value. The tunneling magnetoresistance (TMR) ratio calculated from these simulations is approximately 54.5\%, given by TMR = $(R_{\rm AP}-R_{\rm P})/R_{\rm P}$. The MTJ conductance can be theoretically described by:
\begin{equation} G=G_0[1-P_1P_2\cos(\theta)] \end{equation}
where $G_0$ is the baseline conductance of the junction, and $P_1$ and $P_2$ represent the spin polarizations of the free and fixed layers, respectively. The obtained simulation results confirm the reliability and consistency of our approach with previously reported experimental data.

\begin{figure}[h]
	\centering
        \includegraphics[width=6.5in]{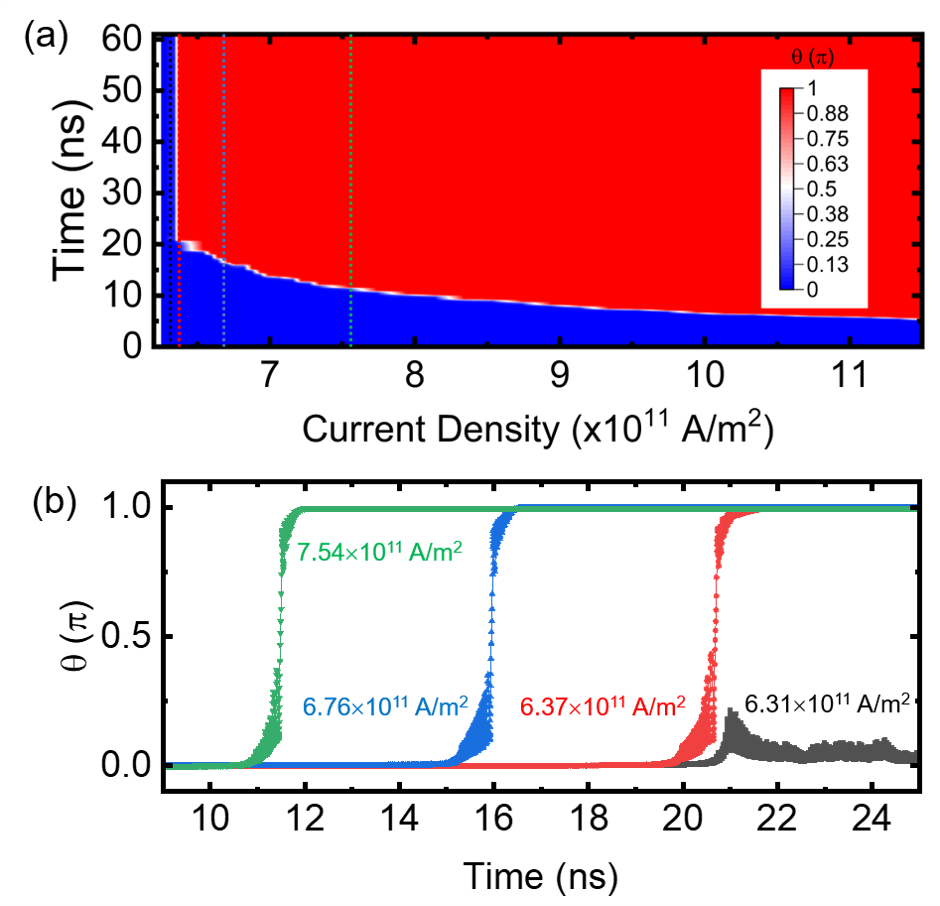}	
	\caption{(a) The color image of $\theta$ as the function of time and current density. (b) The time dependences of the $\theta$ for the MTJ free layer at various current densities.}
 \label{fig2}
\end{figure}

Subsequently, the spin-transfer torque (STT)-induced switching behavior of the free layer was investigated under varying pulse current densities without thermal noise. The applied current pulses had a duration of 20 ns, with total simulation duration set to 100 ns. To better illustrate the switching dynamics, Fig.\ref{fig2}(a) presents a color map of the magnetization angle $\theta$ versus current density and time. At low current densities, such as $6.31\times 10^{11}\ \rm A/m^2$, the magnetization fails to switch in Fig.\ref{fig2}(b). Prior to full magnetization reversal, a narrow region with rapid precessional dynamics appears. With increased current density, the reversal time markedly decreases, achieving a minimum reversal time of about 5 ns. These simulation results closely align with the dynamic characteristics observed experimentally in STT-MRAM devices.

The above simulation results were obtained without considering thermal noise. We further investigated the influence of thermal noise on magnetization switching in MTJs by introducing thermal fluctuations modeled as a white-noise effective magnetic field into the LLG equation. The effective thermal noise field can be expressed as follows \cite{PhysRev.130.1677}:
\begin{equation} 
H=\sqrt{\frac{4K_B T \alpha f}{\gamma M_s H_K}} \end{equation}
where $K_B$ is the Boltzmann constant, $T$ is the absolute temperature, and $f$ is the attempt frequency. 

\begin{figure}[h]
	\centering
        \includegraphics[width=6.5in]{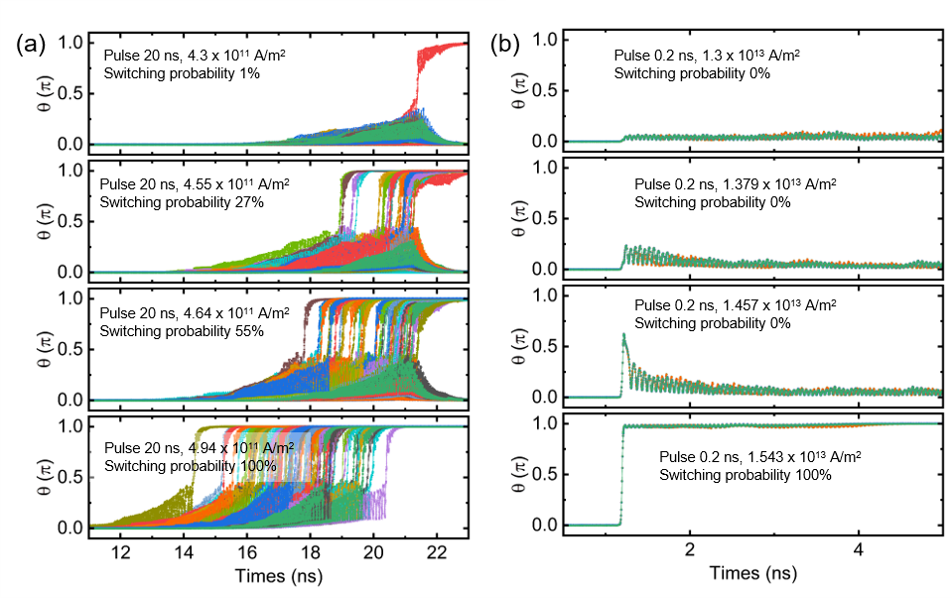}	
	\caption{(a) The time dependence of magnetization angle is calculated and plotted under various current densities with the pulse time of 20 ns by considering the thermal noise under 100 times simulation. (b) The time dependence of magnetization angle is calculated and plotted under various current densities with the pulse time of 0.2 ns by considering the thermal noise under 100 times simulation.  }
 \label{fig3}
\end{figure}

To assess the impact of thermal fluctuations, the temporal evolution of the magnetization angle under various current densities at a pulse duration of 20 ns is presented in Fig.~\ref{fig3}(a).
Notably, the applied current densities are significantly below the deterministic switching current density of $6.37\times 10^{11}$ $\rm A/m^2$. However, probabilistic switching begins to occur at a current density of $4.3\times 10^{11}$ $\rm A/m^2$, with a 1\% switching probability, and achieves 100\% switching probability at a current density of $4.94\times 10^{11}$ $\rm A/m^2$. Conversely, when the pulse duration is shortened to 0.2 ns, probabilistic switching behavior is no longer observed, regardless of increased current density, as demonstrated in Fig.~\ref{fig3}(b). This outcome highlights that current density primarily governs magnetization reversal under short pulse durations, with thermal noise exerting negligible influence

\begin{figure}[h]
	\centering
        \includegraphics[width=6in]{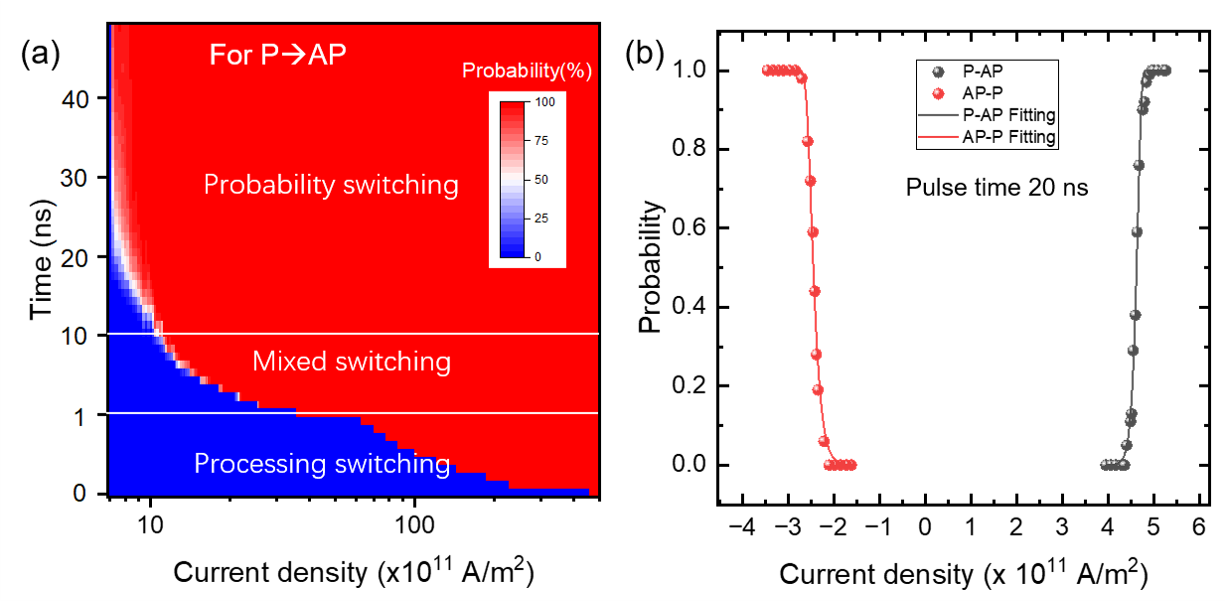}	
	\caption{(a) The relationship between the flip probability of the magnetic moment in the free layer of MTJ and the current and pulse time under the action of thermal noise. (b) When the pulse time is 20ns, the change of MTJ free layer flipping probability under different current densities.  The red curve indicates that the initial state of MTJ is an anti-parallel state (AP), while the black curve indicates that the initial state of MTJ is a parallel state (P). The solid dots are simulation results. The solid lines are the fitting results by heuristic equation.}
 \label{fig4}
\end{figure}

To elucidate the distinct mechanisms involved, Fig.~\ref{fig4}(a) provides a contour plot illustrating the switching probability as a function of current density and pulse duration. For pulse durations exceeding approximately 10 ns, a pronounced transitional region (white region) emerges between $0\%$ and $100\%$ switching probabilities, clearly indicating the dominant role of thermal noise in probabilistic magnetization reversal. In contrast, at pulse durations shorter than 1 ns, the switching probability exhibits an abrupt transition directly from $0\%$ to $100\%$ with increasing current density. This switching behavior aligns well with theoretical predictions of deterministic magnetization switching processes in MTJs.
Additionally, within intermediate pulse durations (approximately 1–10 ns), the probability transition from $0\%$ to $100\%$ is abrupt yet influenced by thermal fluctuations, reflecting a combined effect of current density and thermal noise. Such simulation outcomes are consistent with experimental observations, providing a reliable foundation for designing novel functional devices exploiting the probabilistic switching characteristics of MTJs.

Based on the above results, we have established a clear understanding of the magnetization switching behavior of MTJ devices under thermal noise conditions. Subsequently, we calculated the switching probabilities at varying current densities for different initial magnetization states, as depicted in Fig.\ref{fig4}(b). When the MTJ is initially in a parallel state, applying a positive current switches the magnetization to the antiparallel state. In this scenario, the switching probability begins to rise at a current density of $4.36\times{10}^{11} \mathrm{A/m^2}$, reaching 100\% at approximately $4.49\times{10}^{11} \mathrm{A/m^2}$. Conversely, when the initial state is antiparallel, magnetization switching initiated by a negative current starts at a lower current density of $1.98\times{10}^{11} \mathrm{A/m^2}$, achieving a 100\% probability near $2.68\times{10}^{11} \mathrm{A/m^2}$. This notable asymmetry in switching currents and transition widths arises primarily from heat generation due to spin-current scattering within the tunneling barrier\cite{liu2021important}.

The calculated switching probability aligns well with the heuristic model described by the following equation~\cite{koch2004time,wang2012thermal}:
\begin{equation} 
   P = 1-exp \left\{-\frac{t}{\tau}exp[-\Delta(1-I_{sw}/I_C)] \right \}
\end{equation}
where $t$ represents the duration of the applied switching current pulse (20 ns in our simulations), $\Delta$ denotes the magnetic memory energy barrier, which is influenced by heat arising from spin scattering in the tunneling layer, and $\tau$ is the inverse of the attempt frequency. The critical switching current $I_C$ corresponds to the minimum current required to switch the MTJ resistance within the specified pulse width $t$. These results demonstrate that our simulation model provides an accurate and efficient method for capturing MTJ switching dynamics under the influence of thermal noise.

\section{Conclusion}
In this study, we systematically analyzed the impact of thermal noise on probabilistic switching behavior in STT-MTJ devices using a spin-circuit simulation model. Our findings demonstrate that thermal noise significantly influences magnetization switching, especially for pulse durations exceeding approximately 10 ns, resulting in pronounced probabilistic switching characteristics. Conversely, for pulses shorter than 1 ns, deterministic precessional switching dominates, with minimal thermal influence observed. Within the intermediate pulse-duration regime (1–10 ns), switching behavior exhibits a combination of abrupt and probabilistic transitions, indicating partial thermal-noise contributions. These outcomes validate the accuracy and effectiveness of our spin-circuit model for capturing realistic MTJ switching dynamics under thermal fluctuations. Moreover, the developed spin-circuit simulation framework provides a reliable basis for future MTJ-based circuit designs, including applications such as true random number generators and neuromorphic computing systems, thereby offering valuable insights into advanced spintronic technologies.
%In this study, we have investigated the impact of thermal noise on the probability switching characteristics of STT-MTJs using a spin-circuit simulation model. Our results reveal that thermal noise plays a significant role in the switching behavior of STT-MTJs, particularly for pulse durations longer than 10 ns, where it notably alters the switching probability. For pulses shorter than 1 ns, precessional switching prevents any observable probability switching. In the intermediate range of 1 to 10 ns, a combination of mixed probability and abrupt switching indicates a partial influence of thermal noise on the system. These findings validate the effectiveness of our spin-circuit model in capturing the switching dynamics of MTJs under thermal noise. Additionally, the spin-circuit module developed in this work provides a robust foundation for the design of future circuit systems utilizing MTJ devices, such as true random number generators and neural network computing, offering valuable insights for the development of advanced spintronic applications.

\begin{acknowledgments}
This work is partially supported by Shenzhen Science and Technology Program (JCYJ20240813113228037), Natural Science Foundation of Top Talent of SZTU (GDRC202323), Key Discipline Construction Project of Guangdong Province (2022ZDJS110), Shenzhen Fundamental Research Fund (JCYJ20220818102214030), Key Colleges and Universities Project of Guangdong Province (2023ZDZX1019). 
\end{acknowledgments}

\bibliographystyle{unsrt}
\bibliography{Bib}

\end{document}